# Simple model for efficient search of high-mobility organic semiconductors

Andrey Yu. Sosorev[1,2,3]

1. Faculty of Physics and International Laser Center, Lomonosov Moscow State University, Leninskie Gory 1/62, Moscow 119991, Russia
2. Institute of Spectroscopy of the Russian Academy of Sciences, Fizicheskaya Str., 5, Troitsk, Moscow 108840, Russia
3. Shemyakin-Ovchinnikov Institute of bioorganic chemistry of the Russian Academy of Sciences, Ulitsa Miklukho-Maklaya, 16/10, Moscow, GSP-7, 117997, Russia

## Abstract

High charge mobility in active layers of organic electronic devices is often necessary for their efficient operation. As a result, search for high-mobility materials among the plethora of synthesizable organic semiconductors is of paramount importance for organic electronics. However, a model for rapid but reliable prediction of charge mobility in various organic semiconductors is still lacking. To solve this issue, we propose a simple analytical model that considers the most essential factors governing the charge transport in these materials: intermolecular electronic coupling, charge delocalization, electron-phonon interaction, and static and dynamic disorder. The suggested model efficiently predicts charge mobility for organic semiconductors of various chemical structures and sizes, and significantly outperforms another approach usually used for screening of organic semiconductors, - Marcus model, - which overlooks high-mobility materials possessing small conjugated cores. We thus anticipate that the suggested model is an efficient tool for the search of the high-mobility organic semiconductors, and its possible integration with crystal structure prediction can boost the development of organic electronics.

## Keywords

## 1. Introduction

Organic electronics provides an opportunity for production of (opto)electronic devices that can outperform their inorganic counterparts in many applications. For instance, light-emitting diodes, light-emitting transistors and lasers with desired emission wavelength,[1] sensors with high specificity to the external stimuli,[2,3] photodiodes and organic solar cells with various absorption spectra and photovoltaic properties[4] can be designed. The tunability of the organic electronic devices' properties is provided by the ability to choose the organic semiconductors (OSs) for their active layers from a plethora of synthesizable compounds, which is virtually unlimited.[5,6] In most cases, efficient operation of the organic electronic devices, e.g., organic field-effect transistors (OFETs), requires OSs with high charge mobility, $\mu$, and hence search of the high-mobility OSs is of paramount importance for the area of organic electronics.

For rational choice of the high-mobility OSs from various materials with appropriate functional properties, one needs a tool for their rapid screening.[7] The problem is that rigorous $\mu$ calculation for an OS is highly challenging due to the presence of several factors that affect the charge transport to the comparable extent.[8,9] In contrast to the inorganic semiconductors, where charge carriers are delocalized and band model describes the charge transport rather well, in OSs charge carriers are localized at one or several molecules because of the strong electron-phonon interaction.[10,11,12] In the case of full charge localization at a single molecule, charge hopping models[10,12,13] are appropriate for $\mu$ assessment. However, for high-mobility OSs, especially for organic semiconducting crystals (OSCs) – an OS type showing the highest $\mu$ values[2] – these models are hardly applicable since charge carriers are partially delocalized over several molecules in these materials.[11,12] This follows from the Hall effect[14,15] and decrease of $\mu$ with temperature[15,16] observed in several OSCs and electron spin resonance (ESR) data indicating charge delocalization over ~10 molecules in pentacene.[17] The charge delocalization could be accounted using the band model; however, the latter does not consider charge localization, and hence do not properly describe relative values of $\mu$ and usually significantly overestimates charge mobility.[18,19] Thus, considering the interplay between the charge localization and delocalization is necessary for adequate description of charge transport and $\mu$ estimation in high-mobility OSCs. The most state-of-the-art models of charge transport in OSCs accounting for this interplay are polaron theory,[20,21,22] transient localization scenario[8,23] and delocalization-incorporated nuclear tunneling model (time-dependent wavepacket diffusion) model.[24] However, all the three models are highly computationally demanding and appear to be hardly appropriate for rapid OSs screening.

In this context, hopping model of charge transport has been being a workhorse for assessment of $\mu$ in OSs for many years (see e.g. Refs. [25, 26, 27]). One of its most popular variants is based on the Marcus formula for the rate of charge transfer between the two molecules (sites):[12,13,28]

$$k_M = \frac{2\pi}{\hbar} J^2 \left(\frac{1}{4\pi\lambda kT}\right)^{1/2} \exp\left(-\frac{(\Delta E - \lambda)^2}{4\lambda kT}\right), \qquad (1)$$

where $\hbar$ is the reduced Planck constant, $k$ is the Boltzmann constant, $T$ is the absolute temperature, $J$ is the charge transfer integral describing electronic coupling between the sites, $\lambda$ is the reorganization energy of the site that describes local electron-phonon interaction, and $\Delta E$ is the electron energy difference between the initial and final sites ($\Delta E$=0 if the molecules are identical). According to Eq. (1), the main factors affecting $k_M$ for identical molecules at a given temperature are $J$ and $\lambda$: the larger the former and the lower the latter, the higher the $k_M$, and hence the higher the $\mu$. Being initially formulated for electron transfer in electrolytes,[28] Eq. (1) actually describes the small polaron hopping in OSs as well[12,13,29] and was proven to reasonably estimate the charge transfer rates in many OSCs.[10] Specifically, this simple approach often allows to describe qualitatively the difference in charge mobilities within the series of structurally similar compounds,[30] and was recently proposed as a means for screening the high-mobility OSs.[25-27]

However, several concerns arise with the charge transport model based on Eq. (1) (we will henceforth refer to it as to Marcus model). First, it neglects the abovementioned (coherent) contribution to the charge transport from delocalized charges in the high-mobility OSCs and hence underestimates charge mobility in these materials.[19] This discrepancy is most prominent for OSCs comprising very small molecules, e.g. derivatives of tetracyanoquinodimethane (TCNQ)[31] and tetrathiafulvalene (TTF),[32] which have large $\lambda$ because of strong geometry relaxation following the charge transfer.[33,34] Specifically, the $\lambda$ value exceeds ~250 meV for TTF[34] and TCNQ derivatives,[35,36] which is ca. 2 times larger than that for larger-molecule high-mobility OSCs like pentacene ($\lambda$~100 meV)[30] and rubrene (140 meV).[2] Thus, small-molecule compounds should show low $\mu$ according to the Marcus model even if $J$ values are considerable, since $\lambda$ is in the exponent of Eq. (1). For this reason, for a long time the search for high-mobility OSCs was focused on the larger molecules,[9,37] resulting in overlooking of several excellent OSCs like $F_2$-TCNQ and HM-TTF. Nevertheless, $F_2$-TCNQ shows very high $\mu$ = 7 cm$^2$/(V·s) at room temperature and bandlike charge transport with unusually large negative d$\mu$/d$T$ resulting in $\mu$~25 cm$^2$/(V·s) at 180K.[31] Similarly, $\mu$ values up to 11 cm$^2$/(V·s) were observed for HM-TTF.[32,38] These high $\mu$ can be reconciled with large $\lambda$ if intermolecular charge delocalization, which suppresses geometry relaxation and hence lowers $\lambda$,[36,39,40] is taken into account. Hence, in order not to miss the high-mobility small-molecule OSCs, the impact of charge delocalization on $\lambda$ is to be considered when performing OS screening. Indeed, the model for searching the high-mobility OSCs should be applicable to the materials with very different properties: size, molecular structure, crystal packing motif, and hence different degree of charge delocalization and charge transport mechanism – hopping or coherent. Second, Marcus model predicts $\mu$ increase with temperature (d$\mu$/d$T$>0), in contrast to the bandlike (d$\mu$/d$T$<0) temperature dependence of $\mu$ observed for several OSCs (see e.g. Refs. [16, 31]). This is because Eq. (1) implies thermally activated charge transport, for which thermal fluctuations are beneficial, while in the case of bandlike (coherent) charge transport, thermal motion is detrimental since it disrupts intermolecular charge delocalization.[11,22] Third, the hopping model neglects the role of dynamic disorder, i.e. thermal fluctuations of $J$, which is now widely assumed to limit charge transport in high-mobility OSs.[8,22,23]

To summarize, the main challenge for OSCs screening is to consider the interplay between the charge localization and delocalization in a reasonable but not cumbersome way. Very recently, attempts to address this issue were performed in Refs. [19, 41] yielding models describing the charge transport in wide range of material parameters. In Ref. [42], a phenomenological model assuming the presence of both delocalized and localized states, occupation of which changes with temperature, was shown to accurately reproduce the $\mu(T)$ dependence for tetramethyl–tetraselena–fulvalene (TM-TSF). Attempts to incorporate the charge delocalization in Marcus model were performed in Refs. [36, 43] via dividing $\lambda$ in Eq. (1) by a number of molecules over which the charge is delocalized, $n$. However, neither of the mentioned models did suggested a means for determination of the charge delocalization degree from the material properties, hindering their application to assessment of charge mobility in various OSs.

In this study, we suggest a simple model for charge mobility estimation and screening of high-mobility OSCs. The approach is based on the Marcus formula for charge transfer rate modified to account for the intermolecular charge delocalization, static and dynamic disorder. The key point of the model is the analytical expression for charge delocalization degree as a function of intrinsic material parameters (intermolecular electronic coupling and electron-phonon interaction), disorder and temperature. The proposed model is shown to describe well the experimental $\mu$ data for OSCs with different molecular structure, including that with very small conjugated cores. Thus, the presented approach is anticipated to be an efficient tool for screening of the OSCs to reveal the high-mobility ones.

## 2. Model formulation

### 2.1. Delocalization degree

As mentioned above, the model for efficient screening of high-mobility OSCs requires accounting for the interplay between the intermolecular charge delocalization (facilitated by the strong electronic coupling between the molecules) and charge localization (facilitated by the reorganization, static and dynamic disorder). Thus, an expression for charge delocalization degree is in the heart of such model. We will first obtain it for one-dimensional (1D) case and then extrapolate it to 3D case.

**One-dimensional case**. Assume an infinite 1D lattice. A charge carrier (electron or hole) can be either localized at one site or delocalized over several lattice sites (molecules). Its wavefunction can be written as $\psi = \sum_{i=1}^{N} c_i \varphi_i$, where $\varphi_i$ are the wavefunctions of the charge carrier localized at $i$-th site, and $c_i$ are the corresponding coefficients. The sites' energies, $\varepsilon_i$, are similar in ideal crystal. If the charge carrier occupies one of the sites, the energy of this site decreases due to its geometry relaxation – a polaron is formed. This decrease amounts the polaron binding energy, $E_p = \lambda/2$. Following Ref. [39], we assume that if the charge carrier is delocalized, the site energy decreases by $c_i^4 \cdot \lambda/2$ (proportional to the squared extra charge at the site). The electrons at neighboring sites interact via the charge transfer integrals, $J$; we assume all $J<0$. Within the already mentioned assumptions, the energy of the charge carrier is:

$$E = \sum_{i=1}^{N} \varepsilon_i c_i^2 + \sum_{\substack{j=1 \\ j \neq i}}^{N} \sum_{i=1}^{N} J_{ij} c_i c_j - \frac{1}{2} \sum_{i=1}^{N} c_i^4 \lambda \qquad (2)$$

For simplicity, we will consider the states in which charge carrier is homogeneously distributed over $n$ sites (in reality, charge density at inner sites should be larger[40]), and refer to $n$ as to the charge delocalization degree. The wavefunction of such state reads $\psi = \sum_{i=1}^{n} \frac{1}{\sqrt{n}} \varphi_i$. The reorganization energy in this case decreases in $n$ times as shown schematically in Fig. 1, and Eq. (2) transforms to:

$$E = \varepsilon - \frac{2|J|(n-1)}{n} - \frac{\lambda}{2n}. \tag{3}$$

Eq. (3) yields $E$ as a monotonic function of $n$ as shown in Supporting information (SI), Fig. S1a: $E$ either increases with $n$ when $4J<\lambda$ or decreases with $n$ when $4J>\lambda$. In the former case, charge is localized (Fig. 1a) because of strong electron-phonon interaction, and in the latter case, charge is delocalized (Fig. 1b,c) since intermolecular electronic coupling overwhelms the electron-phonon interaction.

Above, the crystal was considered ideal with equal $\varepsilon_i = \varepsilon$ and $J_i = J$. However, in real crystals, $\varepsilon_i$ suffer the static disorder resulting from the impurities or other crystal defects; the $\varepsilon_i$ standard deviation, $\sigma_s$, is dictated by the crystal quality. Analogously, the $J$ values suffer the dynamic disorder induced mainly by low-frequency intermolecular vibrations;[8,23] the $J$ standard deviation, $\sigma_d$, is governed by the lattice distortion energy (describing non-local electron-phonon interaction), $L$, and $T$: $\sigma_d = \sqrt{2LkT}$.[11,33] Both the static and dynamic disorder facilitate charge localization. To account for this, we employ the concept of potential well analogy[44] and treat the disorder as a decrease in the energy of the localized state. The stronger the disorder and the more localized the state, the deeper the well for it. We set the depth of the well, $\Delta E$, for the carrier localized at single site equal to the standard deviation of the effective site energy, $\sigma$, which includes the static and dynamic contribution: $\sigma = \sqrt{\sigma_s^2 + 2LkT}$ (we assume normal distribution of the site energies and independence of the static and dynamic disorder, resulting in summation of their variances).

Charge delocalization results in "averaging" of the occupied sites' energies and lowers their standard deviation according to the central limit theorem, yielding $\Delta E(n) = \sigma(n) = \frac{\sqrt{\sigma_s^2 + 2LkT}}{\sqrt{n}}$. Thus, the energy of the charge carrier homogeneously distributed over $n$ sites ($n \geq 1$) is

$$E(n) = \varepsilon - \frac{2|J|(n-1)}{n} - \frac{\lambda}{2n} - \frac{\sqrt{\sigma_s^2 + 2LkT}}{\sqrt{n}}, \tag{4}$$

The $E(n)$ function described by Eq. (4) always has a minimum at certain $n=n_{min}$ (see Fig. S1b), in contrast to Eq. (3) (see Fig. S1a). If $4J>\lambda$ and disorder is not too strong, the minimum is achieved for $n_{min}>1$, i.e. charge is delocalized, while if $\lambda >> J$, $E(n)$ function has a minimum at $n_{min}=1$, i.e. charge is localized. Assuming $n>>1$, minimization of Eq. (4) yields $n_{min} = \frac{(2J-\lambda/2)^2}{\sigma_s^2 + 2LkT}$. We extrapolate this expression to lower $n$ as:

$$n_{min} \approx 1 + \frac{(2J-\lambda/2)^2}{\sigma_s^2 + 2LkT} \cdot h(2J-\lambda/2), \tag{5}$$

where $h(x)$ is the Heaviside function.

Eq. (5) provides an analytical expression for assessment of the charge delocalization degree – the key parameter required for description of charge transport in high-mobility OSCs. Albeit being extremely approximate, this expression allows to consider several important aspects of charge transport in high-mobility OSCs: intermolecular charge delocalization and critical role of static and dynamical disorder. Fig.

2a presents $n$ as a function of static disorder and temperature. This figure reveals a peak for low $T$ and $\sigma_S$. The prominent $n$ decrease with $T$ in the case of low $\sigma_S$ explains $\mu$ decrease with $T$ (bandlike transport) typically observed in the high-mobility OSCs, while for large $\sigma_S$, $n$ is low and nearly temperature-independent.

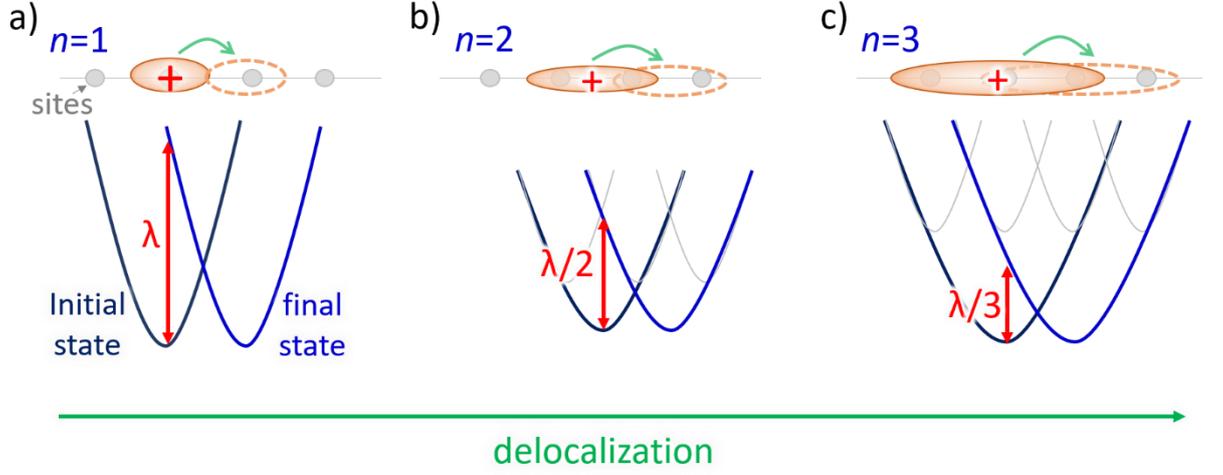

Fig. 1. Illustration of the impact of charge delocalization on the polaron motion within the suggested model.

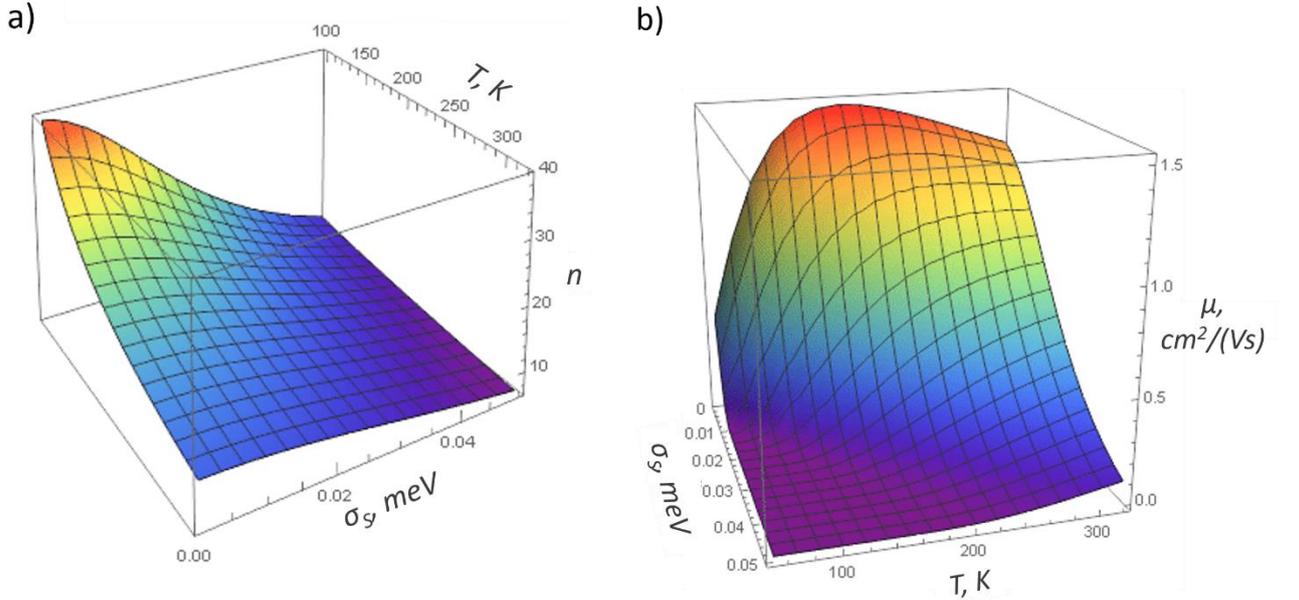

Fig. 2. Charge delocalization degree (a) and charge mobility (b) as a function of temperature and static disorder for $L$ = 25 meV, $J_\Sigma$ = 900 meV and $\lambda$=250 meV.

**Three-dimensional case.** In 3D case, multiple charge transport directions characterized by various transfer integrals, $J_i$, can exist. In Ref. [36], it was shown that the increase of charge transport dimensionality increases the probability of charge delocalization. To account for this fact, we substitute $2|J|$ in Eq. (5) by the sum of absolute $J_i$ values, $J_\Sigma = \sum_{i=1}^{k} |J_i|$, and Eq. (5) transforms to:

$$n_{\min} \approx 1 + \frac{(J_\Sigma - \lambda/2)^2}{\sigma_s^2 + 2LkT} \cdot h(J_\Sigma - \lambda/2) \tag{6}$$

Eq. (6), which describes the charge delocalization degree in an OSC, will be used below for $\mu$ estimation.

## 2.2. Charge transfer rate and charge mobility

In the case of fully localized charge ($n_{min}=1$), the rate of the charge transfer – hopping of the polaron to the neighboring site as sketched in Fig. 1a – can be described by the abovementioned Marcus formula, Eq. (1), which for identical sites ($\Delta E=0$) reduces to

$$k_M = \frac{J^2}{\hbar} \left(\frac{\pi}{\lambda kT}\right)^{1/2} \exp\left(-\frac{\lambda}{4kT}\right) \qquad (7)$$

If the polaron is delocalized, $n_{min}>1$, we treat its motion as the shift of the set of the occupied sites by one site as illustrated in Fig. 1b,c. In this case, we consider that Eq. (7) holds, but the reorganization energy is to be divided by $n_{min}$ in accordance with Refs. [36, 39, 43]. Thus, the transfer rate for (partially) delocalized charge increases as shown in Fig. S2. We also account for $\mu$ dependence on the disorder according to Refs. [10, 45]: $\mu \sim \exp\left(-\left(\frac{2\sigma}{3kT}\right)^2\right)$. Under these assumptions, the charge transfer rate is:

$$k_i = \frac{J_i^2}{\hbar} \left(\frac{\pi n}{\lambda kT}\right)^{1/2} \exp\left(-\frac{\lambda}{4nkT}\right) \exp\left(-\frac{4\left(\sigma_s^2 + 2LkT\right)}{(3kT)^2}\right) \qquad (8)$$

The charge mobility is then calculated using the common Einstein–Smoluchowski relation:

$$\mu = \frac{eD}{kT} = \frac{e}{6kT} \sum_i k_i r_i^2 p_i, \qquad (9)$$

where $D$ is the charge diffusion coefficient, $r_i$ is the distance between the adjacent molecules along the $i$-th transport direction, and $p_i = \frac{k_i}{\sum_j k_j}$ is the probability of the charge to move in this direction.

Eqs. (6,8,9) constitute the suggested approach for the $\mu$ estimation. Fig. 2b plots $\mu$ as a function of $\sigma_S$ and $T$. One can see that in dependence of $\sigma_S$, the model predicts qualitatively different $\mu(T)$ behavior. For strong static disorder (large $\sigma_S$), thermally activated transport is observed. On the contrary, for weak disorder (small $\sigma_S$), bandlike transport takes place. Thus, the model qualitatively reproduces the both types of $\mu(T)$ dependence observed in experiment.

To test the model performance in estimation of the room-temperature $\mu$, we applied it to various OSCs possessing different molecular structures and sizes. The results of the test are described in the next Section.

## 3. Test of the model
### 3.1. Computational

Reorganization energy $\lambda$ was approximated by its inner-sphere part, $\lambda_i$, which is typically considered much larger than the outer-sphere part, $\lambda_o$.[45] The $\lambda$ values were calculated according to the 4-point scheme[11] from the energies of the molecule in 4 states: neutral state in its optimized geometry ($E_N$), neutral state in the optimized geometry of the charged state ($E_N^*$), charged state in its optimized geometry ($E_C$), and charged state in the geometry of the neutral state ($E_C^*$). The energy difference between the former two states, $\lambda_1 = E_N^* - E_N$, describes the energy relaxation of the molecule that have lost the charge carrier, while the energy difference between the latter two states, $\lambda_2 = E_C^* - E_C$, describes the energy relaxation of the molecule that have accepted the charge. Total reorganization energy is $\lambda = \lambda_1 + \lambda_2 = (E_N^* - E_N) + (E_C^* - E_C)$. Transfer integrals $J_i$ were calculated using home-written code based on dimer projection method (DIPRO).[46-48] All DFT calculations were performed using B3LYP functional and 6-31g(d) basis set in GAMESS package.[49,50] Crystal structures were taken from the X-Ray diffraction data in Cambridge Crystallographic Database.

### 3.2. Charge mobilities in various crystalline organic semiconductors

The model was applied to 20 OSCs from the five popular families: (hetero)acenes, oligoarenes, TCNQ derivatives (TCNQs), TTF derivatives (TTFs) and naphthalene-diimides (NDIs). The chemical structures of these compounds are shown in Fig. 3 and Fig. S3.

Oligoacenes (e.g. pentacene), heteroacenes (e.g. dinaphthathienothiophene, DNTT) and their derivatives (e.g. rubrene), are among the most studied OSCs. Their purified samples generally show high charge mobilities $\mu$>1 cm$^2$/(V·s) due to the rigidity of their molecules resulting in small $\lambda$, and extended π-conjugation favorable for large $J$.[45] High $\mu$ in crystalline (hetero)acenes are reproduced even with the Marcus theory,[25,30] although absolute values are underestimated in about 2 times.[25] This discrepancy should be attributed to the neglect of charge delocalization by the Marcus model, while e.g. one of the most studied oligoacene derivatives, rubrene, shows bandlike charge transport[16] and Hall effect[14] – the features that are widely considered as the signatures of delocalized charges.

Oligoarenes, e.g. oligothiophenes, oligophenyls and their co-oligomers, are also among the most popular OSCs due to their synthetical flexibility, high solubility and favorable optoelectronic properties. However, they show low $\mu$, which can be attributed to the flexibility of their molecules resulting in high $\lambda$, and weak intermolecular interactions resulting in low $J$. Marcus theory predicts $\mu$ in these materials reasonably well (see e.g. Ref. [51]), which is natural since no intermolecular charge delocalization is expected in them.

In contrast to the two OSC families mentioned above, for TCNQ, TTF and their derivatives, Marcus model often significantly underestimates the $\mu$ magnitude.[36] These molecules have small conjugated cores as compared to the majority of the high-mobility OSCs, resulting in high $\lambda$ and hence low $\mu$ within the Marcus model (see below). Nevertheless, in experiment they exhibit considerable $\mu$ values up to 11 cm$^2$/Vs for TTFs[32] and 7 cm$^2$/(V·s) for TCNQs[31] as mentioned above. The last family of the OSCs studied, NDI derivatives, have recently become one of the most popular n-type OSCs [52]. For these materials, which possess $\mu$ values up to 7.5 cm$^2$/(V·s),[53] assessment of $\mu$ using Marcus model was not reported. Since several OSCs from the TTF, TCNQ and NDI families show very high $\mu$>5 cm$^2$/(V·s), correct estimation of $\mu$ in this type of materials is necessary for efficient search of high-mobilty OSCs.

Fig. 3 collates the calculated $\lambda$ and $J$ values for several studied OSCs; those for the other investigated materials are presented in Fig. S3. These values are in line with the results of previous calculations for the mentioned OSCs (see, e.g., Refs. [2, 31, 33, 54]). From Figs. 3 and S3 it follows that large $J$>100 meV are observed mostly for oligoacenes and NDI derivatives. This also corresponds to the previous results[31,54] and can be attributed to the extended conjugated cores consisting of condensed aromatic rings, which can provide π-stacking with efficient overlapping of neighboring molecules' orbitals. For (hetero)acenes, the $J$ values clearly increase with the molecular length from tetracene to DATT (cf. Figs 3 and S3). Unexpectedly, HM-TTF that possesses small conjugated core also shows very high $J$>100 meV, which can be attributed to π-stacking with favorable relative position of the molecules in crystal. In the studied crystals of TCNQ derivatives, $J$ do not exceed 70 meV (despite of π-stacking observed in F$_2$-TCNQ), and in the oligoarenes they are below 50 meV because of the herringbone packing motif observed in these materials. Reorganization energies are the largest for NDI derivatives and the lowest for oligoacenes. As a result, within the Marcus theory, $\mu$ values are the highest for oligoacenes (mostly exceeding 1 cm$^2$/Vs), lower for TTF and TCNQ derivatives (0.2-0.8 cm$^2$/Vs), and the lowest ones for NDI derivatives (~0.1 cm$^2$/Vs) and oligoarenes (<0.1 cm$^2$/Vs) as shown below. However, this is in contradiction with the abovementioned high $\mu$>5 cm$^2$/Vs for F$_2$-TCNQ, HM-TTF and NDI-Chex, and points on the necessity for accounting intermolecular charge delocalization.

Within the model suggested in this study (Eqs. (6,8,9)), the delocalization length $n_{min}$ and hence $\mu$ are most affected by the quantity $\Delta = \sum_i |J_i| - \lambda/2 = J_\Sigma - \lambda/2$. If $\Delta$ is positive, charge delocalization ($n_{min}$>1) occurs, otherwise charge localization ($n_{min}$=1) takes place. Indeed, Fig. 3 and S3 show that for OSCs showing high $\mu$ (e.g. rubrene, NDI-CHex, HM-TTF and F$_2$-TCNQ), $J_\Sigma \gg \lambda/2$, while for that showing low $\mu$ (e.g. 4P, 4T, BEDT-TTF), $J_\Sigma < \lambda/2$. Noteworthily, for F$_2$-TCNQ, HM-TTF and NDI-CHex, $\lambda$ exceeds 250 meV – rather high value nearly incompatible with high $\mu$ from the viewpoint of the Marcus theory; nevertheless, $J_\Sigma > \lambda/2$ observed for them enables charge delocalization and high $\mu$ within our model. Importantly, for F$_2$-TCNQ, $J_i$ values along particular directions are not very large (<70 meV), however, 3D character of charge transport resulting in many directions with considerable $J_i$ enables large $\Delta$.

Fig. 4 plots the $\mu$ values calculated using the suggested model and the Marcus one against the experimental ones. Only experimental data obtained using organic field-effect transistors (OFETs) were considered. Static and dynamic disorder were assumed to be similar for all the OSCs, $L$ being set to 50 meV (an estimate based on typical values obtained for oligoacenes[33]), and $\sigma_s$ set to 1 meV (weak static disorder corresponding to highly purified materials; for such a low $\sigma_s$, its impact on $\mu$ is insignificant). Fig. 4 clearly shows that Marcus model significantly underestimates $\mu$ for high-mobility OSCs, the most dramatic discrepancy being observed for TCNQs, TTFs and NDIs: these points are far from the line corresponding to the correlation between the theory and experiment. As mentioned above, the reason for this discrepancy is high $\lambda$ (see Fig. 3) that stems from the small size of the conjugated core and reduces the charge mobility within the Marcus model dramatically. On the contrary, our model predicts reasonable values for all the studied OSCs including small-molecular compounds (TCNQs, TTFs and NDIs) and enables good correlation with the experiment. Specifically, for high-mobility OSCs, e.g. F$_2$-TCNQ and rubrene, the suggested model predicts higher $\mu$ than the Marcus model (and much closer to the experiment) since it considers charge delocalization in these materials. Note that improved correlation with the experiment for our model is observed with $L$ and $\sigma_s$ values equal for various compounds, indicating that the suggested model outperforms the Marcus one not because of the extra adjustable parameters, but due to the consideration of the factors essential for charge transport. It is however worth noting that the model underestimates the $\mu$ values for low-mobility OSCs, e.g. oligoarenes. A possible reason is that the model accounts only detrimental impact of the dynamic disorder, while variation of the $J$ values can be favorable for hopping of localized carriers.[33] In addition, for materials with small $J$, $L$ should be small as well since it is related to $J$ variance, while in Fig. 4 it was taken similar for all the OSCs studied resulting in possible overestimation of the dynamic disorder in low-mobility OSCs.

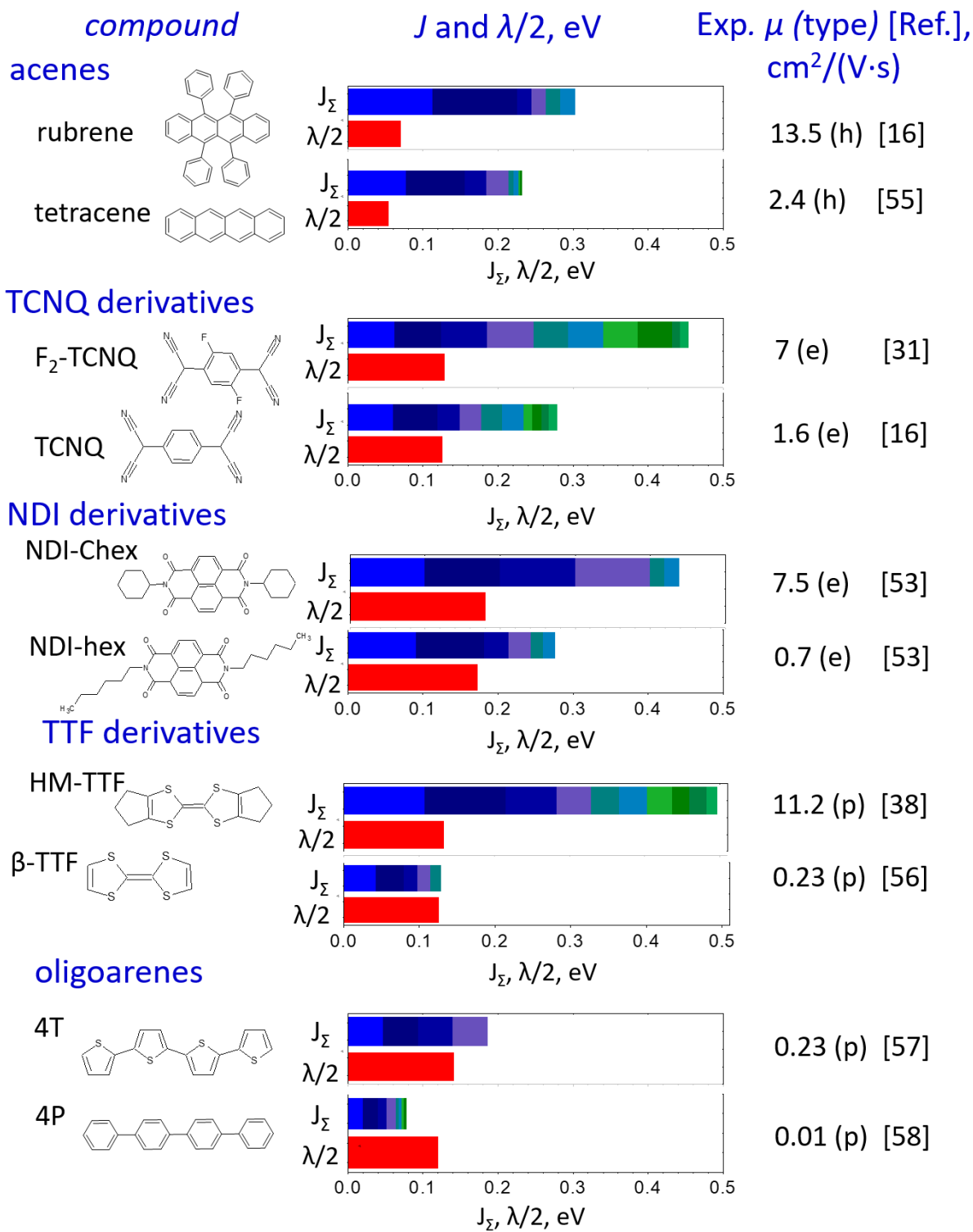

Fig. 3. The calculated transfer integrals over various charge transfer directions, the sums of their absolute values, $J_\Sigma$, and reorganization energies for selected OSCs. $J_\Sigma$ bars are divided to show the $|J_i|$ values along different charge transport directions.

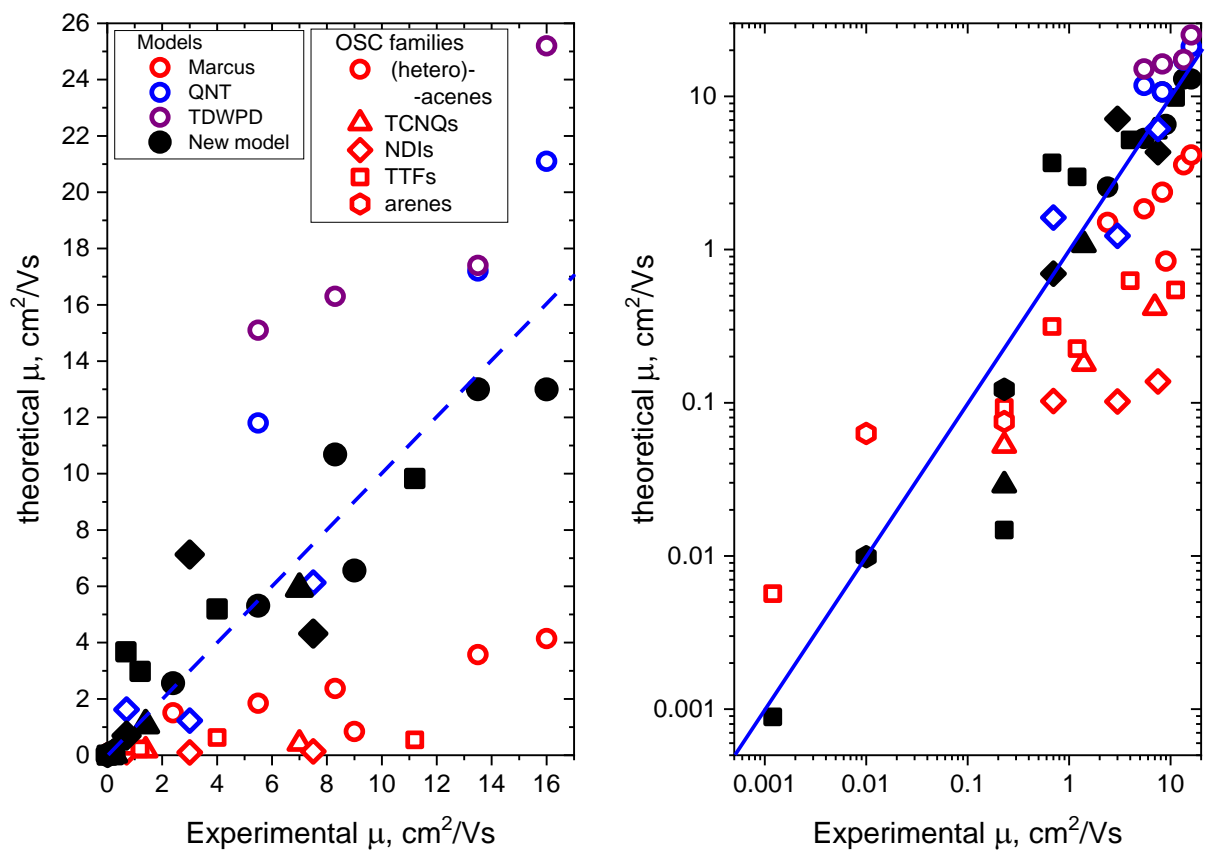

Fig. 4. Comparison of the theoretical $\mu$ values (obtained using Marcus model, red, and the suggested one, black) with the experimental data in linear (a) and logarithmic (b) scales. $L$=50 meV, $\sigma_s$=1 meV. Blue line corresponds to the one-to-one correlation. The experimental OFET $\mu$ data are from Refs. [7, 16, 31, 37, 38, 41, 53, 55-64], see Figs. 3 and S3 for details.

As follows from Fig. 4, the suggested model, albeit being very approximate, can efficiently predict charge mobility in OSCs possessing different conjugated cores and sizes. This opportunity is enabled by the accounting for the interplay between the charge localization and delocalization in OSCs, which is essential for description of the charge transport and prediction of $\mu$. The suggested model is ideologically close to that formulated in Ref. [41], which also considers charge delocalization and static disorder. The $\mu(T)$ dependences obtained in Ref. [41] are close to that observed within our model and also show either bandlike or thermally activated behavior in dependence of the material parameters. However, in contrast to the mentioned model, our approach provides a means for $\mu$ *prediction*, not just for its description. This is enabled by the suggested formula for the degree of charge delocalization – the number of molecules (sites) over which charge carrier is delocalized under various conditions, Eq. (6). Such a predictive power makes our model an efficient tool for the search of high-mobility OSCs. Importantly, if reliable tool for OS crystal structure prediction will be developed, its combination with the suggested model could enable assessment of $\mu$ from just the OS molecular structure – the long-standing challenge in organic electronics. Solution of this problem will provide an unprecedent freedom in design of OSs with the desired functional properties that can boost the development of organic electronics.

## 4. Conclusions

An analytical model capable to predict charge mobility in crystalline organic semiconductors with various molecular structures and sizes is proposed. The heart of the model is the proposed expression for charge delocalization degree as a function of electronic coupling between the molecules, electron-phonon interaction, static and dynamic disorder. The predicted charge mobilities correspond well to the experimental ones. Importantly, the model reproduces high experimental charge mobilities for several OSCs build of small molecules, in contrast to the widely used Marcus model that significantly underestimates them. We thus anticipate that the suggested approach will serve as an efficient tool for screening of the high-mobility OSCs, which are necessary for improved operation of organic electronic devices.

## Conflicts of interest

There are no conflicts of interest to declare.

## Acknowledgements

The work on model formulation was supported by the Russian Foundation for Basic Research (projects # 16-32-60204 mol_a_dk, 19-32-60081). All the calculations were supported by Russian Science Foundation (project # 18-72-10165). The author thanks D. Yu. Paraschuk for valuable discussions and E. Osnachev for assistance in performing calculations.

## References

1. J. Gierschner, S. Varghese and S. Y. Park. *Adv. Optical Mater.,* 2016, **4**, 348.
2. O. Ostroverkhova. *Chem. Rev.*, 2016, **116**, 13279.
3. L. Bai, P. Wang, P. Bose, P. Li, R. Zou and Y. Zhao. *ACS Appl. Mater. Interfaces,* 2015, **7**, 5056.
4. S. Li, W. Liu, C.-Z. Li, M. Shi and H. Chen, *Small,* 2017, **13**, 1701120.
5. C. Wang, H. Dong, W. Hu, Y. Liu and D. Zhu. *Chem. Rev.*, 2012, 112, 2208.
6. J. Mei, Y. Diao, A. L. Appleton, L. Fang and Z. Bao, *J. Am. Chem. Soc.*, 2013, **135**, 6724.
7. L. Wang, G. Nan, X. Yang, Q. Peng, Q. Lia and Z. Shuai. *Chem. Soc. Rev.*, 2010, **39**, 423.
8. S. Fratini, D. Mayou and S. Ciuchi. *Adv. Funct. Mater.*, 2016, **26**, 2292.
9. G. Schweicher, Y. Olivier, V. Lemaur and Y. H. Geerts. *Isr. J. Chem.*, 2014, **54**, 595.
10. A. Köhler and H. Bässler, Electronic Processes in Organic Semiconductors: an Introduction. Wiley-VCH, Weinheim, 2015.
11. Y. Li, V. Coropceanu and J.-L. Brédas, in The WSPC Reference on Organic Electronics: Organic Semiconductors, ed. J.-L. Brédas, S. R. Marder, World Scientific, Singapore, 2016, Chapter 7, 193.
12. H. Oberhofer, K. Reuter and J. Blumberger. *Chem. Rev.*, 2017, **117**, 10319.
13. Z. Shuai, H. Geng, W. Xu, Y. Liao and J.-M. André. *Chem. Soc. Rev.*, 2014, **43**, 2662.
14. V. Podzorov, in Organic Field-Effect Transistors, ed. Z. Bao, J. Locklin, CRC Press, Boca Raton, 2007, Chapter 2.
15. N. A. Minder, S. Ono, Z. Chen, A. Facchetti and A. F. Morpurgo. *Adv. Mater*. 2012, **24**, 503.
16. E. Menard, V. Podzorov, S. H. Hur, A. Gaur, M. E. Gershenson and J. A. Rogers. *Adv. Mater*. 2004, **16**, 2097.
17. K. Marumoto, S. Kuroda, T. Takenobu and Y. Iwasa, *Phys. Rev. Lett.*, 2006, **97**, 256603.
18. Y. Tsutsui, G. Schweicher, B. Chattopadhyay, T. Sakurai, J.-B. Arlin, C. Ruzié, A. Aliev, A. Ciesielski, S. Colella, A. R. Kennedy, V. Lemaur, Y. Olivier, R. Hadji, L. Sanguinet, F. Castet, S. Osella, D. Dudenko, D. Beljonne, J. Cornil, P. Samorì, S. Seki and Yves H. Geerts, *Adv. Mater.*, 2016, **28**, 7106.
19. I. Yavuz, *Phys. Chem. Chem. Phys.*, 2017, **19**, 25819.
20. F. Ortmann, F. Bechstedt and K. Hannewald. *Phys. Status Solidi* B, 2011, **248**, 511.
21. I. G. Lang and Y. A. Firsov. *J. Exp. Theor. Phys.*, 1963, **16**, 1301.
22. A. Yu. Sosorev, D. R. Maslennikov, O. G. Kharlanov, I. Yu. Chernyshov, V. V. Bruevich and D. Yu. Paraschuk. *Phys. Status Solidi RRL,* 2018, **13**, 1800485.
23. A. Troisi and G. Orlandi, *Phys. Rev. Lett.,* 2006, **96**, 086601.
24. Y. Jiang, X. Zhong, W. Shi, Q. Peng, H. Geng, Y. Zhao and Z. Shuai, *Nanoscale Horiz.,* 2016, **1**, 53.
25. I. Yavuz, B. N. Martin, J. Park and K. N. Houk. *J. Am. Chem. Soc.,* 2015, **137**, 2856.
26. I. Yavuz, S. A. Lopez, J. B. Linc and K. N. Houk. *J. Mater. Chem. C,* 2016, **4**, 11238.
27. C. Schober, K. Reuter and H. Oberhofer. *J. Phys. Chem. Lett.,* 2016, **7**, 3973.
28. R.A. Marcus and N. Sutin, *Biochim. Biophys. Acta,* 1985, **811**, 265.

29. A. Troisi. *Chem. Soc. Rev.*, 2011, **40**, 2347.
30. W.-Q. Deng and W. A. Goddard III. *J. Phys. Chem. B,* 2004, **108**, 8614.
31. Y. Krupskaya, M. Gibertini, N. Marzari and A. F. Morpurgo. *Adv. Mater.,* 2015, **27**, 2453.
32. R. Pfattner, S. T. Bromley, C. Rovira and M. Mas-Torrent. *Adv. Funct. Mater.,* 2016, **26**, 2256
33. R. S. Sánchez-Carrera, P. Paramonov, G. M. Day, V. Coropceanu and J.-L. Brédas. *J. Am. Chem. Soc.*, 2010, **132**, 14437.
34. H.-X. Li, R.-H. Zheng and Q. Shi. *Phys. Chem. Chem. Phys*., 2011, **13**, 5642.
35. I. Y. Chernyshov, M. V. Vener, E. V. Feldman, D. Y. Paraschuk and A. Y. Sosorev, *J. Phys. Chem. Lett.*, 2017, **8**, 2875.
36. A. Yu. Sosorev. *Phys. Chem. Chem. Phys*., 2017, **19**, 25478.
37. A. N. Sokolov, S. Atahan-Evrenk, R. Mondal, H. B. Akkerman, R. S. Sánchez-Carrera, S. Granados-Focil, J. Schrier, S. C. B. Mannsfeld, A. P. Zoombelt, Z. Bao and A. Aspuru-Guzik. *Nat. Commun*., 2011, **2**, 437.
38. Y. Takahashi, T. Hasegawa, S. Horiuchi, R. Kumai, Y. Tokura and G. Saito, *Chem. Mater.,* 2007, **19**, 6382.
39. S. Larsson and A. Klimkans, *Molecular Crystals and Liquid Crystals,* 2001, **355**, 217.
40. S. T. Bromley, F. Illas and M. Mas-Torrent. *Phys. Chem. Chem. Phys*., 2008, **10**, 121.
41. C. Liu, K. Huang, W.-T. Park, M. Li, T. Yang, X. Liu, L. Liang, T. Minari and Y.-Y. Noh. *Mater. Horiz*., 2017, **4**, 608.
42. H. Xie, H. Alves and A. F. Morpurgo. *Phys. Rev. B,* **80**, 245305.
43. B. Blülle, A. Troisi, R. Häusermann and B. Batlogg. *Phys. Rev. B,* 2016, **93**, 035205.
44. C. M. Soukoulis and E.N. Economou, *Waves Random Media*, 1999, **9,** 255.
45. V. Coropceanu, J. Cornil, D. A. da Silva Filho, Y. Olivier, R. Silbey and J.-L. Brédas. *Chem. Rev.*, 2007, **107**, 926.
46. B. Baumeier, J. Kirkpatrick and D. Andrienko, *Phys. Chem. Chem. Phys.*, 2010, **12**, 11103.
47. J. Kirkpatrick, *Int. J. Quantum Chem.*, 2008, **108**, 51.
48. H. Kobayashi, N. Kobayashi, S. Hosoi, N. Koshitani, D. Murakami, R. Shirasawa, Y. Kudo, D. Hobara, Y. Tokita and M. Itabashi, *J. Chem. Phys.*, 2013, **139**, 8.
49. M. W. Schmidt, K. K. Baldridge, J. A. Boatz, S. T. Elbert, M. S. Gordon, J. H. Jensen, S. Koseki, N. Matsunaga, K. A. Nguyen, S. Su, T. L. Windus, M. Dupuis and J. A. Montgomery, *J. Comput. Chem.* 1993, **14**, 1347.
50. M. S. Gordon and M.W. Schmidt, in *Theory and Applications of Computational Chemistry: the first forty years*, ed. C. E. Dykstra, G. Frenking, K. S. Kim and G. E. Scuseria, Elsevier, Amsterdam, 2005; 1167.
51. X.Yang, L. Wang, C. Wang, W. Long and Z. Shuai, *Chem. Mater.,* 2008, **20**, 3205.
52. K. Zhou, H. Dong, H.-l. Zhang and W. Hu. *Phys. Chem. Chem. Phys*., 2014, **16**, 22448.
53. D. Shukla, S. F. Nelson, D. C. Freeman, M. Rajeswaran, W. G. Ahearn, D. M. Meyer and J. T. Carey. *Chem. Mater.,* 2008, **20**, 7486.
54. S. P. Adiga and D. Shukla. *J. Phys. Chem. C,* 2010, **114**, 2751.
55. C. Reese, W.-J. Chung, M.-M. Ling, M. Roberts and Z. Bao. *Appl. Phys. Lett.,* 2006, **89**, 202108.
56. H. Jiang, X. Yang, Z. Cui, Y. Liu, H. Li, W. Hu, Y. Liu and D. Zhu, *Appl. Phys. Lett.*, 2007, **91**, 123505.
57. C. Reese, M. E. Roberts, S. R. Parkin, and Z. Bao, *Adv. Mater.,* 2009, **21**, 3678.
58. D. J. Gundlach, Y. Y. Lin, T. N. Jackson and D. G. Schlom, *Appl. Phys. Lett.,* 1997**, 71**, 3853.
59. S. Lee, B. Koo, J. Shin, E. Lee, H. Park and H. Kim. *Appl. Phys. Lett*., 2006, **88**, 162109.
60. S. Haas, Y. Takahashi, K. Takimiya and T. Hasegawa. *Appl. Phys. Lett.,* 2009, **95**, 022111.
61. Q. Xin, S. Duhm, F. Bussolotti, K. Akaike, Y. Kubozono, H. Aoki, T. Kosugi, S. Kera and N. Ueno. *Phys. Rev. Lett.,* 2012, **108**, 226401.
62. T. He, Y. Wu, G. D'Avino, E. Schmidt, M. Stolte, J. Cornil, D. Beljonne, P. P. Ruden, F. Würthner and C. D. Frisbie. *Nat. Commun*., 2018, **9**, 2141.
63. S. Tamura, T. Kadoya, T. Kawamoto and T. Mori, *Appl. Phys. Lett.,* **2013**, *102*, 063305.
64. M. Mas-Torrent, P. Hadley, S. T. Bromley, X. Ribas, J. Tarrés, M. Mas, E. Molins, J. Veciana and C. Rovira. *J. Am. Chem. Soc.,* 2004, **126**, 8546.

# Supporting Information

## Simple model for efficient search of high-mobility organic semiconductors
Andrey Yu. Sosorev

### 1. Impact of model parameters

#### Charge delocalization degree

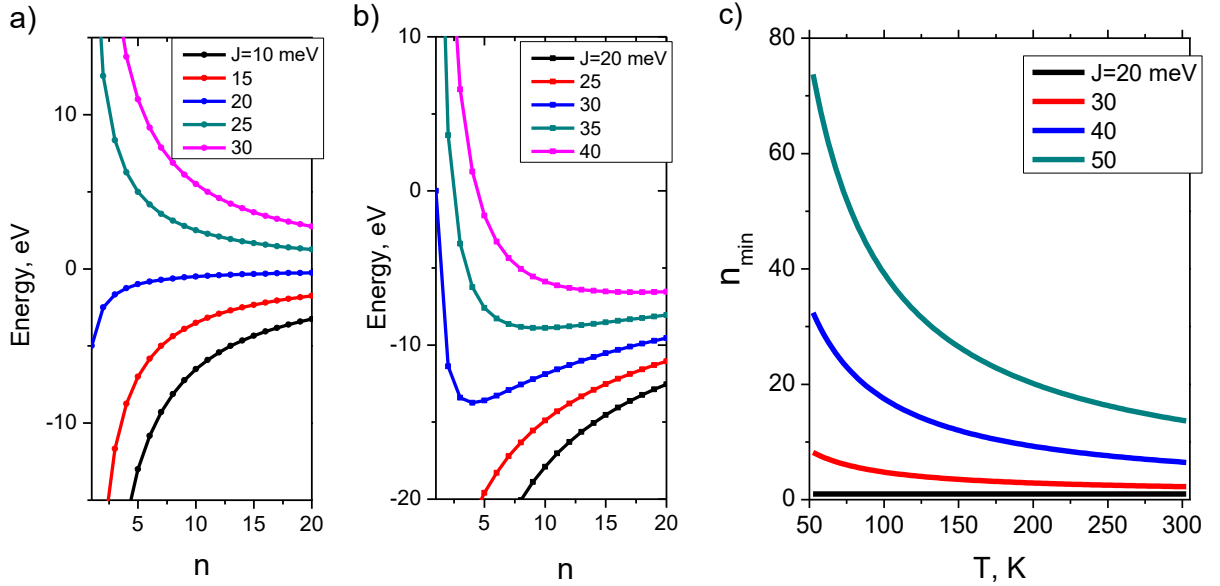

Fig. S1. Charge delocalization degree for different conditions. (a,b) Polaron energy as a function of $n$ for various $J$ values without disorder (a) and with static and dynamic disorder, $L$=50 meV and $σ_S$=25 meV. (c) Delocalization length, $n_{min}$, as a function of temperature, $T$. $λ$=250 meV.

#### Impact of charge delocalization on charge mobility

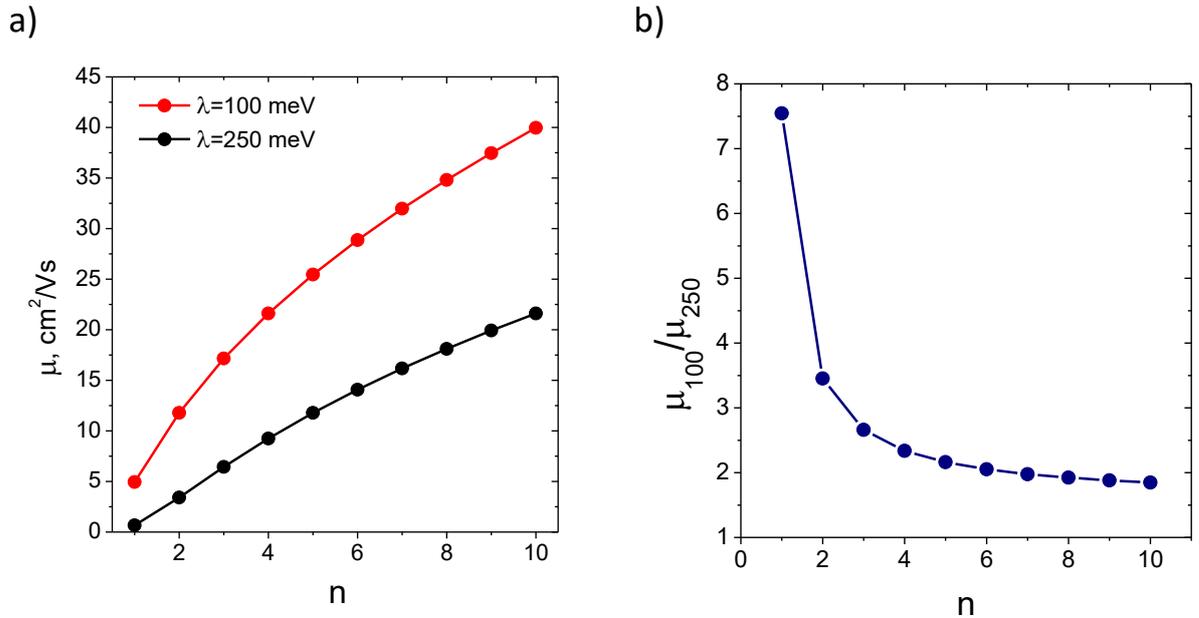

Fig. S2. (a) Increase of the charge mobility with the increase of the charge delocalization degree (number of sites over which electron is delocalized) within the suggested model for $λ$=100 and 250 meV. $J$=70 meV along all the 3 directions, $L$=0, $σ_S$=0. (b) The ratio of the charge mobilities for $λ$=100 and 250 meV as a function of charge delocalization.

2. Various materials

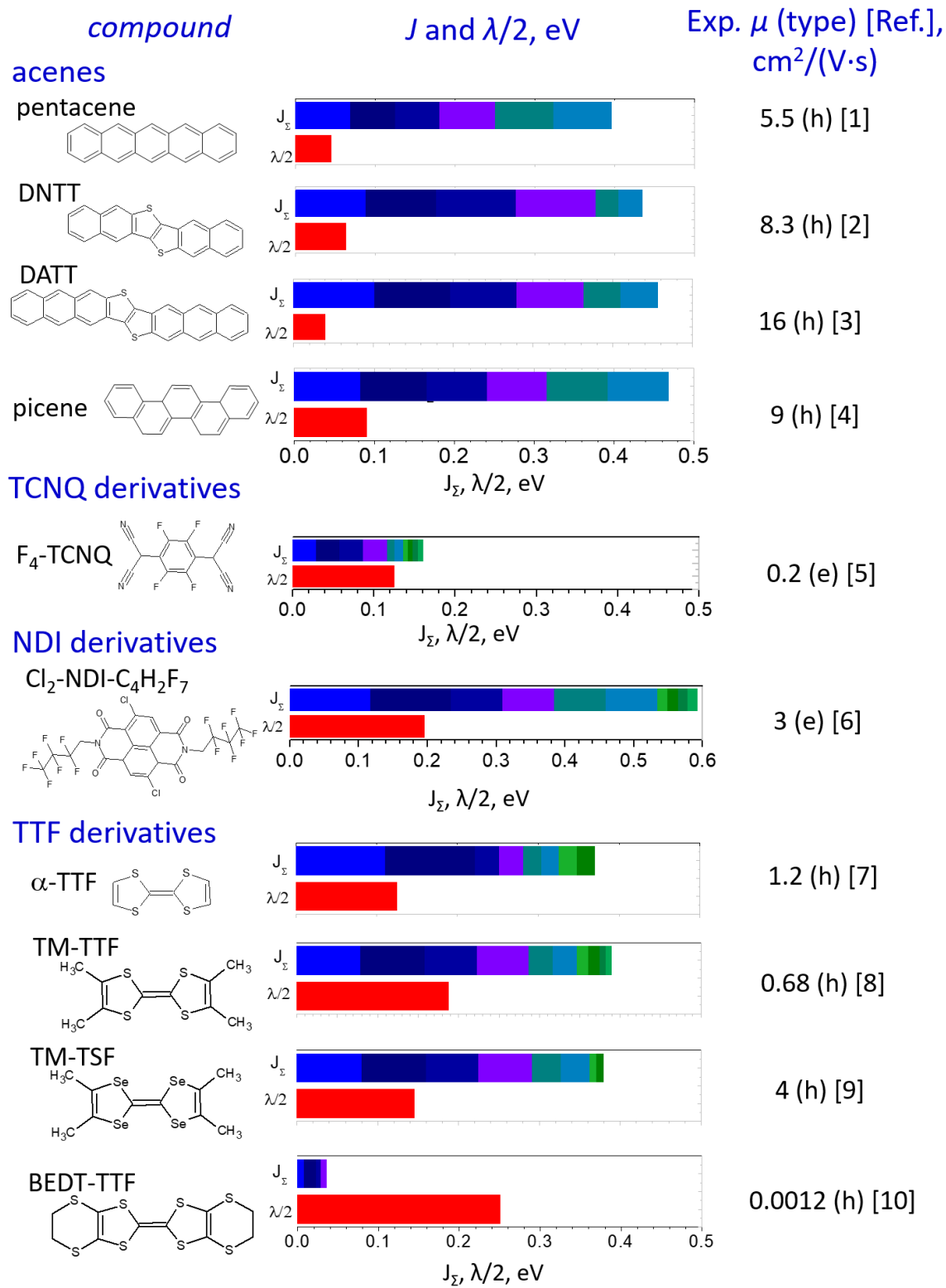

Fig. S3. The calculated transfer integrals over various charge transfer directions, the sums of their absolute values, $J_\Sigma$, and reorganization energies for the OSCs not included in Fig. 3 of the main text. The experimental OFET $\mu$ data are from the Refs. provided at the right side of the Figure. $J_\Sigma$ bars are divided to show the |J| values for different charge transport directions.

## References

1. S. Lee, B. Koo, J. Shin, E. Lee, H. Park and H. Kim. *Appl. Phys. Lett.* 2006, **88**, 162109.
2. S. Haas, Y. Takahashi, K. Takimiya and T. Hasegawa. *Appl. Phys. Lett*. 2009, **95**, 022111.
3. A. N. Sokolov, S. Atahan-Evrenk, R. Mondal, H. B. Akkerman, R. S. Sánchez-Carrera, S. Granados-Focil, J. Schrier, S. C. B. Mannsfeld, A. P. Zoombelt, Z. Bao and A. Aspuru-Guzik. *Nat. Commun.*, 2011, **2**, 437.
4. Q. Xin, S. Duhm, F. Bussolotti, K. Akaike, Y. Kubozono, H. Aoki, T. Kosugi, S. Kera and N. Ueno. *Phys. Rev. Lett*. 2012, **108**, 226401.
5. Y. Krupskaya, M. Gibertini, N. Marzari, A. F. Morpurgo, Adv. Mater., 2015, 27, 2453-2458.
6. T. He, Y. Wu, G. D'Avino, E. Schmidt, M. Stolte, J. Cornil, D. Beljonne, P. P. Ruden, F. Würthner and C. D. Frisbie. *Nat. Commun*., 2018, **9**, 2141
7. H. Jiang, X. Yang, Z. Cui, Y. Liu, H. Li, W. Hu, Y. Liu, D. Zhu, *Appl. Phys. Lett.* 2007, **91**, 123505
8. S. Tamura, T. Kadoya, T. Kawamoto, T. Mori. *Appl. Phys. Lett.* **2013**, *102*, 063305.
9. H. Xie, H. Alves, A. F. Morpurgo, *Phys. Rev. B* **2009**, *80*, 245305.
10. M. Mas-Torrent, P. Hadley, S. T. Bromley, X. Ribas, J. Tarrés, M. Mas, E. Molins, J. Veciana and C. Rovira. *J. Am. Chem. Soc.* 2004, **126**, 8546